\documentclass[twocolumn,aps,showpacs,preprintnumbers,amsmath,amssymb]{revtex4}

\usepackage{graphicx}
\usepackage{dcolumn}
\usepackage{bm}

\begin{document}

\title{Fusion Enhancement for Neutron-Rich Light Nuclei}

\author{Varinderjit Singh}
\author{J. Vadas}
\author{T.~K. Steinbach}
\author{S. Hudan}
\author{R.~T. deSouza}
\email{desouza@indiana.edu}
\affiliation{%
Department of Chemistry and Center for Exploration of Energy and Matter, Indiana University\\
2401 Milo B. Sampson Lane, Bloomington, Indiana 47408 USA}%

\author{L.~T. Baby}
\author{S.~A. Kuvin}
\author{V. Tripathi}
\author{I. Wiedenh\"{o}ver}
\affiliation{
Department of Physics, Florida State University, Tallahassee, Florida, 32306 USA}%

\date{\today}

\begin{abstract}
\begin{description}
\item[Background] Measurement of the fusion cross-section for neutron-rich light nuclei is crucial in ascertaining if fusion of these nuclei occurs in the outer crust of a neutron star.
\item[Purpose] Measure the fusion excitation function at near-barrier energies for the $^{19}$O + $^{12}$C system. Compare the experimental results with the fusion excitation function of
$^{18}$O + $^{12}$C and $^{16}$O + $^{12}$C.
\item[Method] A beam of $^{19}$O, produced via the $^{18}$O(d,p) reaction, was incident on a $^{12}$C target at energies near the Coulomb barrier. Evaporation residues 
produced in fusion of $^{18,19}$O ions with $^{12}$C target nuclei were detected with good geometric efficiency and identified by measuring their
energy and time-of-flight. 
\item[Results] A significant enhancement is observed in the fusion probability of $^{19}$O ions with a $^{12}$C target as compared to $^{18}$O ions. 
\item[Conclusion] The larger cross-sections observed at near barrier energies 
is related to significant narrowing of the fusion barrier indicating a larger tunneling probability for the fusion process. 
\end{description}
\end{abstract}

\maketitle

Approximately half the elements beyond iron are formed via the r-process in which seed nuclei rapidly capture 
multiple neutrons and subsequently undergo $\beta$ decay. Although it is clear that a high neutron density is required for
the r-process, 
the exact site or sites at which r-process nucleosynthesis occurs is still a question of debate. One 
proposed scenario 
involves the merging of two compact objects such as neutron stars. Tidal forces between the two compact objects 
disrupts the neutron stars, ejecting neutron-rich nuclei into the interstellar medium. 
Although nucleosynthesis via decompression of neutronized nuclear matter was 
initially proposed 
decades ago \cite{Lattimer77,Meyer89}, 
only recently have detailed computational investigations of such a scenario e.g. tidal disruption of
a neutron star become feasible 
\cite{Berger13,Martin13,Foucart14,Just15,Radice16}. The most recent calculations suggest that such events could
be responsible for heavy element (A$>$130) r-process nucleosynthesis. 
Recent observation of gravitational waves emanating from two black holes 
merging \cite{Abbott16} has re-ignited the question of whether and to what degree the disruption 
of neutron stars contributes to the heavy element composition of the universe.

A natural question in considering the ejecta from the disruption of the neutron star is the 
composition of the neutron star prior 
to the merger as well as the reactions that might occur both during and post the merging event. 
The outer crust of a neutron star provides an unique environment in which nuclear reactions can occur. 
Of particular interest are the fusion reactions of neutron-rich light nuclei. These nuclei have been hypothesized 
to fuse more readily than the corresponding $\beta$ stable isotopes providing 
a potential heat source that triggers the fusion of $^{12}$C 
nuclei resulting in an X-ray superburst \cite{Horowitz08}. 
An initial measurement of fusion induced with neutron-rich oxygen nuclei suggested 
an enhancement of the fusion probability as compared to standard models of fusion-evaporation \cite{Rudolph12}. 
To definitively establish if neutron-rich light nuclei exhibit a fusion enhancement at sub-barrier energies, high quality 
experimental data is needed. 
In the present work, we present for the first time a measurement of the 
total fusion cross-section for $^{19}$O + $^{12}$C at incident energies near the barrier 
and compare the results with the fusion cross-section for $^{16,18}$O + $^{12}$C.

Fusion excitation functions reflect the interplay of the repulsive 
Coulomb and attractive nuclear potentials as the two nuclei collide. As the charge distribution of the projectile oxygen nuclei is essentially 
unaffected by the additional neutrons, the repulsive Coulomb potential is unchanged. Consequently, the comparison of the fusion excitation functions 
for the different oxygen isotopes provides access to the changes in the attractive nuclear potential. This change in the attractive potential 
can be related to changes in the neutron density distribution with increasing number of neutrons for oxygen nuclei.

\begin{figure}
\includegraphics[scale=0.7]{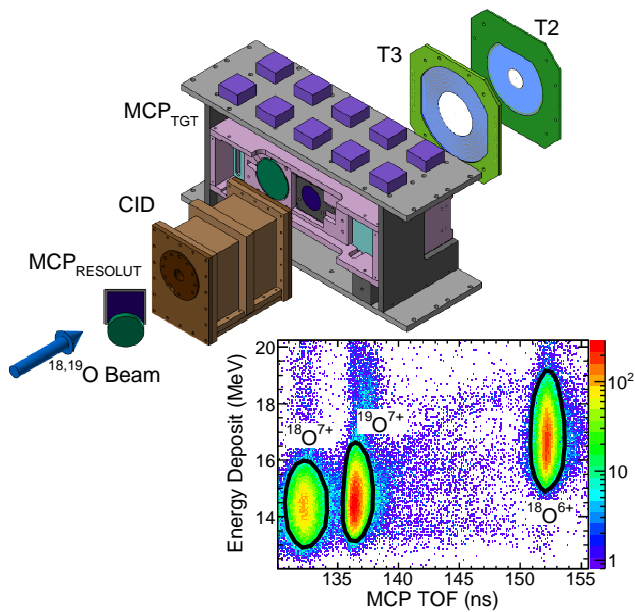}
\caption{\label{fig:setup} (Color online) Schematic illustration of the experimental setup. 
The MCP$\mathrm{_{RESOLUT}}$ detector is located approximately 3.5 m upstream of the MCP$\mathrm{_{TGT}}$ detector.
Inset: Energy deposit versus time-of-flight spectrum for ions exiting RESOLUT that are incident on $^{12}$C target 
at E$_{\mathrm{lab}}$=46.7 MeV.
Color is used to represent yield in the two dimensional spectrum on a logarithmic scale.}
\end{figure}

The experiment was performed at the John D. Fox accelerator laboratory at Florida State University. A beam of $^{18}$O ions, accelerated to an 
energy of 80.7 MeV impinged on a deuterium gas cell at a pressure of 350 torr cooled to a temperature of 77 K. Ions of $^{19}$O were produced via 
a (d,p) reaction and separated from the incident beam by the electromagnetic spectrometer RESOLUT \cite{RESOLUT}. 
Although this spectrometer rejected 
most of the unreacted beam that exited the production gas cell, the beam exiting the spectrometer 
consisted of both $^{19}$O and $^{18}$O ions. 
As each beam particle was independently identified, this beam mixture allowed simultaneous measurement of $^{18}$O + $^{12}$C 
and $^{19}$O + $^{12}$C thus providing a robust measure of the fusion enhancement due to the presence of the additional neutron.
The experimental setup used to measure fusion of oxygen ions with carbon nuclei is depicted in Fig.~\ref{fig:setup}. 
To identify beam particles, the energy deposit and time-of-flight \cite{deSouza11} of each particle was measured. 
Upon exiting the spectrometer particles first traverse a thin secondary emission foil (0.5 $\mu$m thick aluminized mylar) ejecting electrons in the process. 
These electrons are accelerated and bent out of the beam path and onto the surface of a 
microchannel plate detector (MCP$\mathrm{_{RESOLUT}}$) where they are 
amplified to produce a fast timing signal. After traversing the thin foil of MCP$\mathrm{_{RESOLUT}}$, the ions passed through a compact 
ionization detector (CID) located approximately 3.5 m downstream. 
Passage of the ions through this ionization chamber results in an 
energy deposit ($\Delta$E) characterized 
by their atomic number (Z), mass number (A), and incident energy. After exiting the small ionization chamber the ions are incident on a 100 $\mu$g/cm$^2$ carbon foil. This foil serves both as a secondary electron emission foil 
for the target microchannel plate detector (MCP$\mathrm{_{TGT}}$) and as the target for the fusion experiment.

\begin{figure}
\includegraphics[scale=0.4]{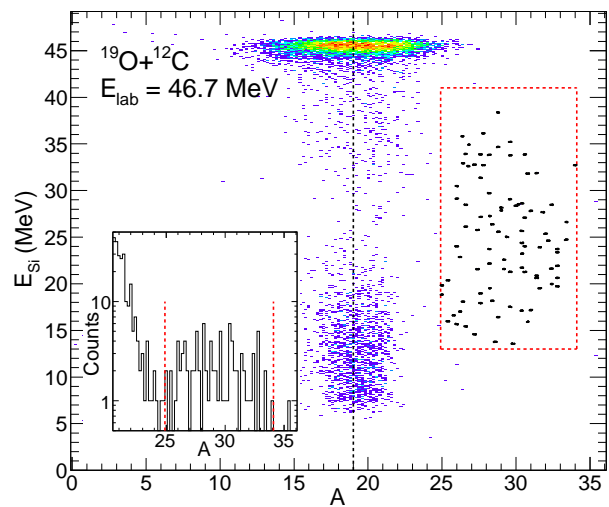}
\caption{\label{fig:pid} (Color online) Two dimensional spectrum depicting dependence of the energy deposited in the annular 
silicon detector, T2, on the mass of the ion. The dashed (red) rectangle indicates the region of the evaporation residues. 
Inset: Mass distribution of ions detected within the interval 13 MeV $<$ E$_{Si}$ $<$ 41 MeV. Vertical lines indicate the A limits 
used to designate evaporation residues.}
\end{figure}

By utilizing the timing signals from both microchannel plate detectors together with the ionization chamber 
a $\Delta$E-TOF measurement is performed. 
This measurement allows identification of ions in the beam as indicated in the inset of Fig.~\ref{fig:setup}. Clearly evident in the figure are three peaks associated with the $^{19}$O$^{7+}$ ions, $^{18}$O$^{7+}$ ions, and $^{18}$O$^{6+}$ ions. 
The $^{19}$O ions corresponded to 31 \% of the beam intensity with the $^{18}$O$^{7+}$ and $^{18}$O$^{6+}$ corresponding 
to approximately 20 \% and 29 \% respectively.
Fusion of $^{19}$O (or $^{18}$O) nuclei in the beam together with $^{12}$C nuclei in the target foil results in the production of an excited $^{31}$Si 
(or correspondingly $^{30}$Si) nucleus. For collisions near the Coulomb barrier the excitation of the fusion product is relatively modest, E$^*$ $\approx$ 35 MeV. 
This fusion product de-excites by evaporation of a few neutrons, protons, and $\alpha$ particles resulting in evaporation residues (ERs). 
Statistical model calculations \cite{evapor} indicate that for $^{31}$Si compound nucleus, the nuclei
$^{30}$Si, $^{29}$Si, $^{28}$Si, $^{29}$Al, $^{28}$Al, $^{27}$Mg, and $^{26}$Mg account for the bulk of the ERs. 
These ERs are deflected from the beam direction by the recoil imparted by the emission of the light particles.
The ERs are detected and identified by two annular silicon detectors designated T2 and T3 
situated downstream of the MCP$\mathrm{_{TGT}}$. These detectors  subtend the angular range 
3.5$^\circ$ $<$ $\theta_{lab}$ $<$ 25$^\circ$. Evaporation residues 
are distinguished from scattered beam, as well as emitted light particles, by measuring their time-of-flight 
between the MCP$\mathrm{_{TGT}}$ detector and the silicon detectors together with 
the energy deposit in the Si detector, E$_{Si}$. Using the measured energy deposit, E$_{Si}$ and the time-of-flight, 
the mass of the ion can be calculated. 
Shown in Fig.~\ref{fig:pid} is the representative two-dimensional mass-energy 
distribution for particles 
incident on the T2 detector at an incident energy in the laboratory of 46.7 MeV for $^{19}$O$^{7+}$ beam. The most prominent feature of this spectrum is the peak at 
E$_{Si}$ = 45 MeV and A=19 which corresponds to elastically scattered beam particles. 
At lower energies than this peak 
with a mass centered on A=19
one also observes a ridge of intensity corresponding to beam particles that are scattered in the 
experimental setup downstream of the target but
prior to entering the silicon detector. 
This scattered beam corresponds to approximately 15 \% of the elastic peak intensity. Situated at 
higher mass number than the scattered beam and with energies 13 MeV $<$ E$_{Si}$ $<$ 41 MeV are detected ions that correspond to evaporation residues. 
The mass distribution associated with this energy interval is presented in the inset of Fig.~\ref{fig:pid}. 
A clear peak in the mass distribution is evident 
at A $<$ 30. The peak is clearly separated from the tail of the scattered beam particles. The centroid and second moment of this peak was determined in 
the interval 24.5 $<$ A   $<$ 34. The measured $\langle$A$\rangle$ for the evaporation residues is 29 and the second moment $\sigma_{ER}$ measured = 2.32. The 
measured width of the evaporation residue mass resolution is largely dictated by the time and energy resolution of the measurement as is evident 
from the width of the elastic peak ($\sigma_{elastic}$ = 2.44).

The fusion cross-section is extracted from the measured yield of evaporation residues through the 
relation $\sigma_{fusion}$ = N$_{ER}$/($\epsilon_{ER}$ x t x N$_{INCIDENT}$) where 
N$_{INCIDENT}$ is the number of beam particles of a given type incident on the target, t is the target 
thickness, $\epsilon_{ER}$ is the detection efficiency, and N$_{ER}$ 
is the number of evaporation residues detected. The number N$_{INCIDENT}$ is determined by counting the 
particles with the appropriate time-of-flight 
between the two microchannel plates that additionally have the correct identification in the $\Delta$E-TOF map 
depicted in the inset of Fig.~\ref{fig:setup}. 
The target thickness, t, of 105 $\mu$g/cm$^2$ is provided by the manufacturer and has an uncertainty 
of $\pm$ 0.5 $\mu$g/cm$^2$. The number of detected residues, 
N$_{ER}$, is determined by summing the number of detected residues with the appropriate mass and energy 
as indicated in Fig.~\ref{fig:pid}. To obtain the detection efficiency, $\epsilon_{ER}$,
a statistical model is 
used to describe the de-excitation of the fusion product together with the geometric acceptance of the experimental setup. 
The detection efficiency varied from 37 \% at the highest incident energies measured to 42 \% at the 
lowest incident energy due to the changing kinematics of the reaction.

\begin{figure}
\includegraphics[scale=0.65]{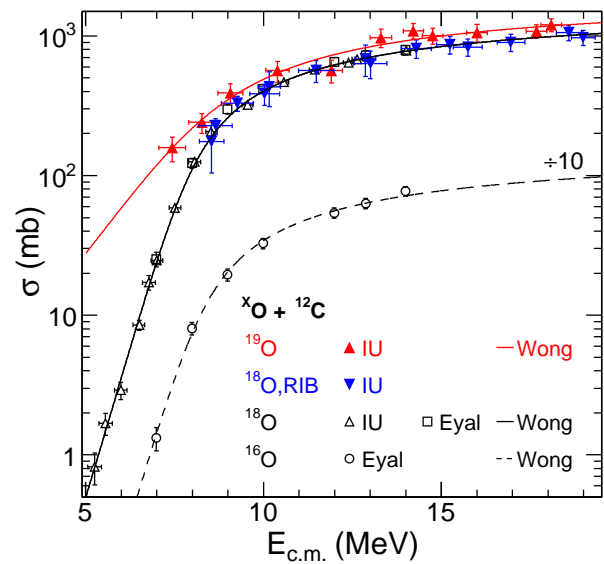}
\caption{\label{fig:xsect} (Color online) 
Fusion excitation function for $^{16,18,19}$O ions incident on $^{12}$C target nuclei. 
The data for $^{16}$O + $^{12}$C have been scaled down by a factor of 10.}
\end{figure}

Presented in Fig.~\ref{fig:xsect} is the dependence of the fusion cross-section on incident energy for
$^{19}$O + $^{12}$C (red triangles) and $^{18}$O + $^{12}$C (blue triangles) measured in the present experiment. Also
shown for comparison is a prior high resolution measurement (open triangles) \cite{Steinbach14a} 
together with older data from the literature \cite{Eyal76} for 
$^{18}$O + $^{12}$C. 
This high resolution measurement utilized a direct high quality beam of $^{18}$O with a similar 
experimental setup to the present experiment \cite{Steinbach14,Steinbach14a}.
It is clear from Fig.~\ref{fig:xsect} that the measured cross-section for 
the $^{18}$O beam in the present experiment (blue points) 
is in good agreement with the previous high resolution measurements (open triangles and squares). This result 
provides confidence in the radioactive beam cross-sections simultaneously measured in the present experiment. 
Also shown in Fig.~\ref{fig:xsect} (open circles) is the fusion excitation function for $^{16}$O + $^{12}$C. 
The data presented utilizes only direct 
measurement of evaporation residues \cite{Eyal76} to characterize the excitation function. 
This excitation function has also been measured by detection of 
$\gamma$ rays in a thick-target experiment. The data from that experiment \cite{Cujec76} is in reasonably 
good agreement with the excitation function depicted. 
However, as use of the thick target measurements is subject to different uncertainties, we omit these data in order to 
make the most straightforward and relevant comparison.

All of the excitation functions depicted in Fig.~\ref{fig:xsect} manifest the same general trend. 
With decreasing incident energy the cross-section decreases as expected for a barrier controlled process.
Closer examination of the $^{19}$O and $^{18}$O reactions reveals that
the $^{19}$O data exhibits a larger fusion cross-section 
as compared to the $^{18}$O data at essentially all energies measured. 
The most important feature of the measured excitation functions is that at the lowest energies measured the fusion 
cross-section for the $^{19}$O system decreases more gradually with decreasing energy than does the $^{18}$O system.
In order to better quantify these differences in the fusion excitation functions 
we have fit the measured cross-sections with a simple 
one dimensional barrier penetration model. The Wong formalism \cite{Wong73} considers the 
penetration of an inverted parabolic barrier with the cross-section given by: 
\begin{equation}
\sigma = \frac{R_C^2}{2E}\hbar\omega \cdot ln \left \{ 1+exp\left [\frac{2\pi}{\hbar\omega}(E-V_C)\right] \right \}
\end{equation} 
where E is the incident energy, V$_C$ is the barrier height, R$_C$ is the radius of interaction and $\hbar$$\omega$ 
is the barrier curvature. 
The fit of the high resolution $^{18}$O data and 
the $^{16}$O data are indicated as the solid black and dashed black lines in Fig.~\ref{fig:xsect} respectively. 
The good agreement observed between the Wong fit of the 
high resolution $^{18}$O data and the $^{18}$O data measured in this experiment (blue points) underscores that 
there are no significant systematic errors associated with the present measurement. 
The solid red curve in Fig.~\ref{fig:xsect} depicts the fit of the $^{19}$O data. 
With the exception of the cross-section measured at E$_{cm}$ $\approx$ 12 MeV, 
the measured cross-sections are reasonably described by the Wong formalism. 
The extracted parameters for the $^{16}$O, $^{18}$O,  and $^{19}$O reactions are summarized in Table 1.
It is not surprising that the barrier height, V$_C$, remains essentially the same for all of the three reactions examined as 
the charge density distribution is unchanged. 

\begin{table}[h]
\centering
\caption{\label{tab:5/tc}Wong fit parameters for the indicated fusion excitation functions. See text for details.}
\begin{ruledtabular}
\begin{tabular}{|c|c|c|c|}

 & V$_C$ (MeV) & R$_C$ (fm) & $\hbar$$\omega$ (MeV) \\
\hline
$^{16}$O + $^{12}$C & 7.93 $\pm$0.16 & 7.25 $\pm$ 0.25 & 2.95 $\pm$ 0.37 \\

$^{18}$O + $^{12}$C & 7.66 $\pm$ 0.10 & 7.39 $\pm$ 0.11 & 2.90 $\pm$ 0.18\\

$^{19}$O + $^{12}$C & 7.73 $\pm$ 0.72 & 8.10 $\pm$ 0.47 & 6.38 $\pm$ 1.00 \\

\end{tabular}
\end{ruledtabular}

\end{table}

\begin{figure}
\includegraphics[scale=0.65]{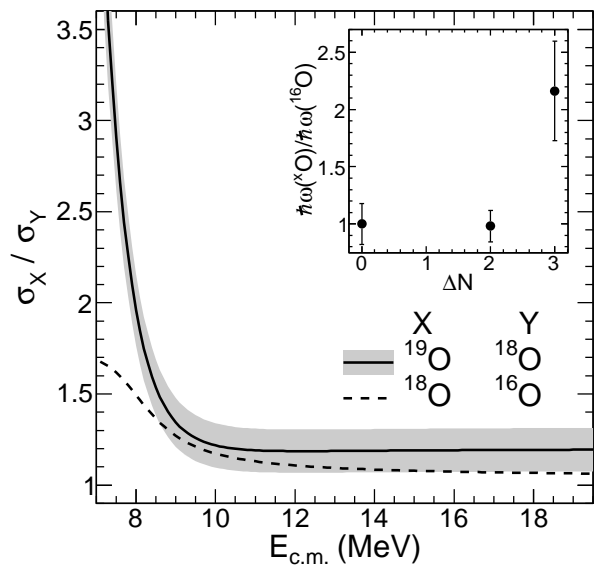}
\caption{\label{fig:xs_ratio} 
Dependence of the ratio of $\sigma$($^{19}$O)/$\sigma$($^{18}$O) and 
$\sigma$($^{18}$O)/$\sigma$($^{16}$O) on E$_{c.m.}$. The shaded region depicts
the uncertainty associated with the ratio for the $^{19}$O reaction. 
Inset: Dependence of the ratio of the barrier curvature on neutron number 
for the $^{19}$O and $^{18}$O reactions as compared to the $^{16}$O reaction. 
}
\end{figure}

The dependence of the ratio  $\sigma$($^{19}$O)/$\sigma$($^{18}$O) and 
$\sigma$($^{18}$O)/$\sigma$($^{16}$O) on E$_{c.m.}$ is shown in Fig.~\ref{fig:xs_ratio}. 
The solid line corresponds the ratio of the Wong fit for $^{19}$O to the Wong fit for $^{18}$O. 
For E$_{c.m.}$ $>$ 10 MeV, 
this ratio is essentially constant at a value of approximately 1.2.
For E$_{c.m.}$ $<$ 10 MeV, the quantity $\sigma$($^{19}$O)/$\sigma$($^{18}$O) increases rapidly reaching a value 
of $\approx$3.0 at the lowest energy measured. It is useful to examine the behavior of 
$\sigma$($^{18}$O)/$\sigma$($^{16}$O) 
presented as the dashed line in Fig.~\ref{fig:xs_ratio}. for comparison.  
At energies well above the barrier $\sigma$($^{18}$O)/$\sigma$($^{16}$O)  is essentially flat at a value of $\approx$1.1.
As one approaches the barrier it increases to a value of approximately 1.7. Hence the 
enhancement observed for $^{19}$O is significantly larger than in the case of $^{18}$O.

In the energy domain above the barrier one expects the ratio of the two cross-sections to be governed by the ratio of 
their geometric cross-sections namely the square of the ratio of their radii.  
The 20\% increase in the cross-section observed for $^{19}$O relative to $^{18}$O can thus be associated with a larger radius for 
$^{19}$O as compared to $^{18}$O within the framework of an inverted parabolic barrier penetration model.
The magnitude of the increase 
in R$_C$, for the $^{19}$O reaction as compared to the $^{18}$O reaction,
is 0.71 fm, 
which corresponds to a 
relative increase of approximately 10\%. This increase is significantly larger than the increase of 0.14 fm for 
$^{18}$O reaction as compared to $^{16}$O. 
This increase in the radius is significantly larger than that expected based upon a 
standard A$^{1/3}$ dependence
emphasizing the fact that in these low energy fusion reactions, the initial interpenetration of the matter 
distributions of the two nuclei is small. Consequently, it is the interaction between the two low-density tails
of the colliding nuclei that governs whether the fusion occurs. As the A$^{1/3}$ dependence does not describe 
the behavior of the low-density tail it is unsurprising that the experimental data deviates from this behavior.

Near and below the barrier, one expects the cross-section to be governed by the detailed shape of the barrier.
Within the context of the Wong formalism this is reflected by the barrier curvature, $\hbar$$\omega$.
Shown in the inset of Fig.~\ref{fig:xs_ratio} is the curvature of the barrier for $^{18,19}$O + $^{12}$C as compared to
$^{16}$O + $^{12}$C as a function of additional neutrons. It is clearly evident that while the additional two neutrons in 
$^{18}$O as compared to $^{16}$O do not substantially alter the barrier curvature, the presence of the additional 
unpaired neutron in $^{19}$O significantly increases the barrier curvature. The barrier in the case of $^{19}$O is a factor of 2.2 
thinner than in the case of $^{16}$O resulting in greater penetration and an enhancement of the fusion cross-section.
Although the success of static models in describing the fusion of stable light nuclei is well established, 
it is unclear whether fusion of neutron-rich nuclei necessitates consideration of collective modes i.e. dynamics as the two nuclei
fuse. Comparison of the present experimental data with more sophisticated models such as a density constrained TDHF 
model \cite{Keser12} are presently underway. 

In summary, we have measured for the first time the fusion of $^{19}$O + $^{12}$C at incident energies near and below the barrier. 
This measurement probes the open question of whether fusion of light neutron-rich nuclei is enhanced relative to their $\beta$ 
stable isotopes.
Comparison of the fusion excitation function for $^{19}$O + $^{12}$C with that of $^{18}$O + $^{12}$C, 
clearly demonstrates that for the $^{19}$O system, fusion is significantly 
enhanced. Well above the barrier this enhancement is approximately 20 \% which can be related to an increase in the 
radius of $\approx$ 10\% due just to the presence of the additional neutron. Near and below the barrier the fusion enhancement 
is even more dramatic, increasing to a factor of three at the lowest energy measured. The dramatic increase in this 
energy domain is related to a significant reduction in the width of the fusion barrier for $^{19}$O as compared to $^{18}$O.
Within the context of an inverted parabolic model the barrier for $^{19}$O is 2.2 times narrower than that for $^{18}$O.
The decrease in the width of the barrier with increasing neutron number suggests that even more neutron-rich oxygen isotopes may
exhibit even narrower barriers and larger fusion enhancements. These results motivate the investigation of even more neutron-rich 
light nuclei, particularly at energies near and below the barrier.

We wish to acknowledge the support of the staff at Florida State University's John D. Fox accelerator in providing the high quality beam that 
made this experiment possible. This work was supported by the U.S. Department of Energy under Grant No. DE-FG02-88ER-40404 (Indiana University), Grant no. DE-FG02-02ER-41220 (Florida State University) and
the National Science Foundation under Grant No PHY-1491574 (Florida State University). J.V. acknowledges the support of a NSF Graduate Research Fellowship
under Grant No. 1342962.

\bibliography{fusion_19O}


\end{document}